\documentclass[pre,aps,amssymb,amsmath,superscriptaddress,twocolumn,showpacs,]{revtex4-1}

\usepackage{graphicx}

\newcommand{\D}{\mathcal{D}}
\newcommand{\e}{\mathbf{e}}
\renewcommand{\epsilon}{\varepsilon}

\newcommand{\kk}{\mathbf{k}}

\newcommand{\p}{\mathbf{p}}
\renewcommand{\phi}{\varphi}

\newcommand{\x}{\mathbf{x}}

\def\be{\begin{equation}}
\def\ee{\end{equation}}
\def\bea{\begin{eqnarray}}
\def\eea{\end{eqnarray}}

\def\media#1{\langle#1\rangle}

\def\xx{{\bf x}}
\def\yy{{\bf y}}

\let\a=\alpha     \let\g=\gamma     \let\d=\delta    
\let\e=\varepsilon
  \let\h=\eta     \let\th=\theta     
\let\l=\lambda
          \let\x=\xi        \let\p=\pi       

\let\s=\sigma \let\t=\tau           
\let\c=\chi
   \let\o=\omega     
 \let\D=\Delta       \let\L=\Lambda

\begin{document}

\title{Height fluctuations in non-integrable classical dimers}

\author{Alessandro Giuliani}
\affiliation{Dipartimento di Matematica e Fisica, Universit\`a degli Studi Roma
Tre, L.go S. L. Murialdo 1, 00146 Roma - Italy}

\author{Vieri Mastropietro}
\affiliation{Dipartimento di Matematica, Universit\`a degli Studi di Milano, 
Via Saldini, 50, 20133 Milano - Italy}

\author{Fabio Lucio Toninelli}
\affiliation{CNRS and Universit\'e de Lyon, Universit\'e Lyon 1, Institut Camille Jordan,
43 bd du 11 novembre 1918,
69622 Villeurbanne - France}

\begin{abstract}
We rigorously establish the asymptotic equivalence between the height function of interacting dimers on the square lattice and the massless Gaussian free field. Our theorem explains the microscopic origin of 
the sine-Gordon field theory description away from the free fermion point, which has previously been elusive. 
We use a novel technique, based on the combination of 
discrete holomorphicity with exact, constructive, renormalization group methods,
which has the potential of being applicable to a variety of other non-integrable models at or close to criticality. 
\end{abstract}


\maketitle

  
High temperature superconductivity and the physics of Resonance Valence Bonds (RVB) \cite{A}
was the original motivation for studying two-dimensional (2D) quantum {\it dimers},
which later became an important model for frustrated magnetism, cold bosons, and many other systems with hard 
constraints \cite{MO}.
In these contexts also classical dimers are of interest, 
not only because they capture the high temperature physics of their quantum counterpart, but also
because for special values of the parameters the quantum static correlations
can be expressed in terms of the classical ones
\cite{RK}. The properties of a wide class of classical dimer models
can be understood by using a celebrated result of half a century ago,
the Kasteleyn theorem \cite{K}, ensuring exact solvability and explicit expressions of the correlations,
which can be written in terms of {\it Pfaffians}. By using this result and the above mentioned equivalence,
the correlations of certain quantum dimer models at special values of the parameters
on the square \cite{RK} and triangular lattice \cite{S} were computed, finding
a power law (critical), and an exponential (massive) large distance decay, respectively.

However, exact solvability 
is limited to a special class of systems, and further progress in our understanding of the physics of dimers requires the analysis 
of  what happens away from integrability.  
We consider 
a prototypical 
non-solvable dimer model
obtained by assuming a local interaction between parallel dimers: given a
periodic box $\L\subset\mathbb Z^2$ of side $L$ (with $L$ even), the partition function is
\be Z_\L(\l,m)=\sum_{M\in \mathcal M_{\L}}\Big[\prod_{b\in M}t^{(m)}_b\Big] e^{\l
\sum_{P\subset\L}N_P(M)}
\label{1}\ee
where
$\mathcal M_{\L}$ is the set of dimer coverings of $\L$, $\l=v/T$ with $T$ the temperature,
$P$ is a plaquette (face of
$\mathbb Z^2$) and $N_P(M)=1$ if the plaquette $P$ is occupied by two
parallel dimers in $M$, and $N_P(M)=0$ otherwise; the $m$-dependence in the reference weight
$t^{(m)}_{(\xx,\xx+\hat e_j)}=1+\d_{j,1}m(-1)^{x_1}$ tunes the distance from criticality; $\l$ tunes the distance from solvability, 
with $\l>0$ corresponding to a local attractive interaction.
This model, in the $m=0$ case, describes polar crystals \cite{O}
and it was recently reconsidered
in \cite{AJMPMT, P,CCMP,D,TSH,RPM}  where its connection with quantum dimer models, RVB physics and large spin quantum anti-ferromagnets was worked out in detail and used to infer informations on the RVB spin-liquid order parameters. MonteCarlo simulations show the presence of
{\it non-universal} anomalous exponents in the dimer correlations decay. 
This confirms the general picture that 
the asymptotic properties can be captured by a {\it quantum field theory} (QFT) of the sine-Gordon type, the fundamental field being 
a coarse-grained version of the {\it height} function.
Using this effective QFT description, several informations were derived about the phase diagram, including the 
Kosterlitz-Thouless universality of the phase transition from a liquid to a crystalline phase. 
The same effective description is believed to be applicable to a variety of dimer and interface models, and it is at the basis of our current understanding of their physics.
However,
while the validity of the QFT description is supported {\it a posteriori} 
by the agreement of  its prediction with simulations, a 
purely deductive and rigorous microscopic argument establishing its correctness 
is currently not available \cite{MO}, with the only exception of the integrable, non-interacting, case. Even then,
the derivation is very non-trivial, and it has been provided only recently \cite{Ke}
using Discrete Holomorphicity (DH) methods.

In this letter we present the first mathematical justification of the quantum field theory description of
{\it non-integrable} dimer models. We prove a theorem establishing the convergence, in the scaling limit, of the height function of  model (\ref{1})
to the massless Gaussian Free Field (GFF), in a suitable range of parameters.
This is done by a
new method, based on the combination of
DH methods  with Constructive Renormalization Group (CRG) techniques \cite{M}, which  
can be applied in a much wider context, including interacting dimers on different lattices and non-integrable deformations of Ising models. 

Given a dimer covering $M$, two faces of $\L$ 
centered at $\xx$ and $\yy$ and a path  $C_{\xx\to \yy}$ from $\xx$ to
$\yy$ with trivial winding around the torus $\Lambda$, we define the {\it height} difference between $\xx$ and $\yy$ as 
\be 
h_\xx-h_{\yy}=\sum_{b\in \mathcal C_{\xx\to \yy}}\big({\openone}_b(M)-\frac14\big)\s_b\label{1.2}\ee
where $\s_b=+1/-1$ depending on whether $C_{\xx\to \yy}$ 
crosses $b$ with the white site on the right/left. 
\begin{figure}[ht]
\includegraphics[width=.2
\textwidth]{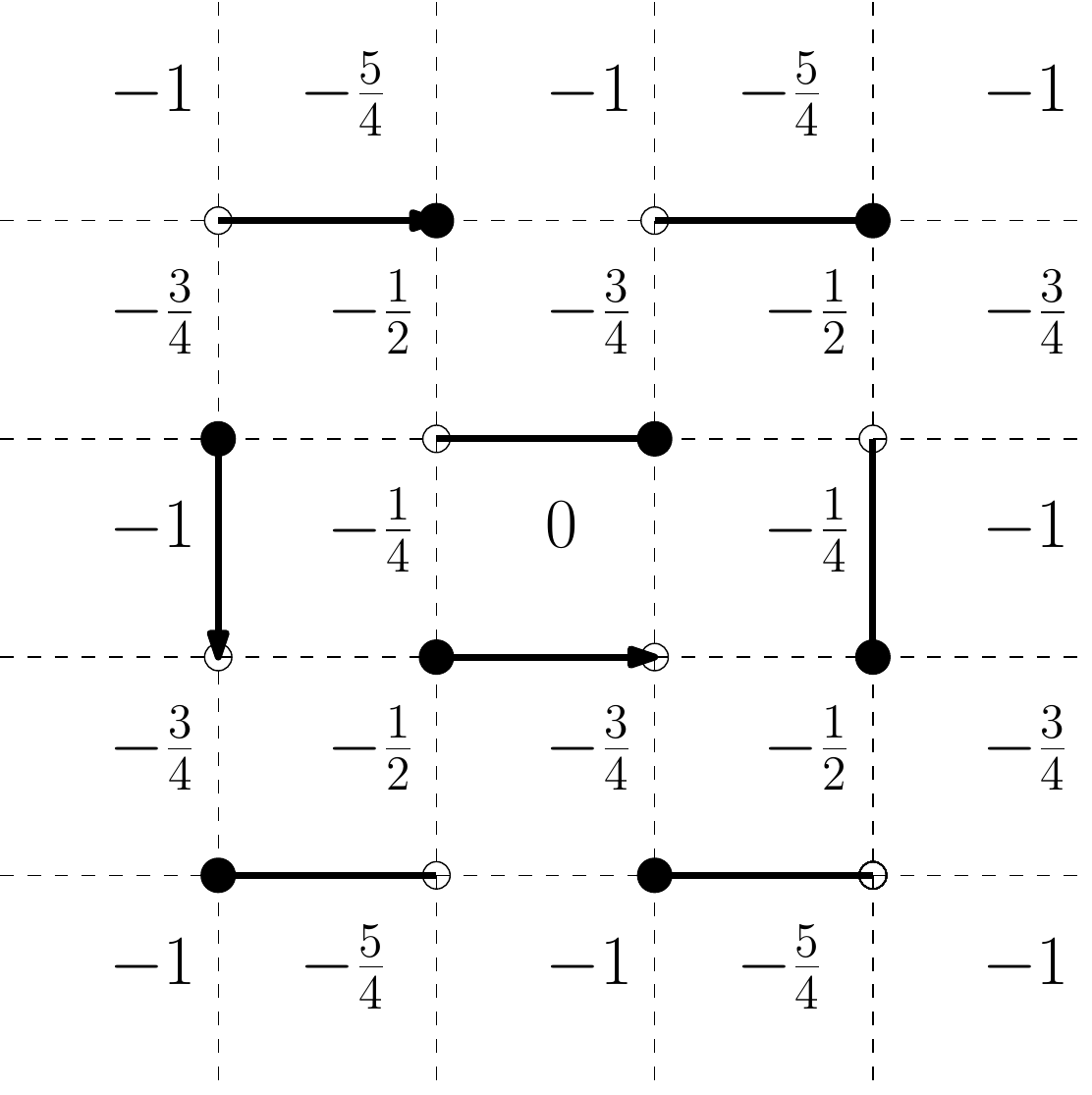}
\caption{A dimer configuration for $L=4$ and the associated height function. The height of the central plaquette is conventionally set to 0.
}
\label{figaltezza}
\end{figure}
Moreover, ${\openone}_b(M)$ is equal to 1 if $b$ is occupied by a dimer
in $M$, and 0 otherwise. 
A crucial property of the height function is
that $h_\xx-h_{\yy}$ is {\it independent} of the choice of
$C_{\xx\to \yy}$.
The {\it dimer correlation} is given by 
$\media{{\openone}_b;{\openone}_{b'}}$, where $\media{\dots;\dots}$ is the truncated expectation with the weight in (\ref{1}),
and the two point 
{\it 
height correlation} is
\be \langle\,(h_{\xx}-h_{\yy})^2\,\rangle=\sum_{b_1\in \mathcal C_{\xx\to \yy}} \sum_{b_2\in \mathcal C_{\xx\to \yy}}\s_{b_1}\s_{b_2}\media{{\openone}_{b_1};
{\openone}_{b_2}}.\label{e31}\ee
Our main result is the following.
\vskip.2truecm
{\bf Theorem.} {\it For $\l\neq 0$ sufficiently small, $L\to\infty$, and $m\to 0$, 
the height correlation for $\xx\neq\yy$ verifies:
\be  \media{(h_\xx-h_{\yy})^2}=\frac{K(\l)}{\p^2}\log|\xx-\yy|+R(\xx-\yy),\label{e1.1}\ee
with $K(\l)$ an analytic function such that $K(0)=1$, and 
$R(\xx)$ a bounded remainder. The higher order truncated correlations are bounded uniformly in $|\xx-\yy|$. At large distances, the coarse graining of $h_\xx$ converges to the Gaussian Free Field (GFF), in the sense that, if $\alpha\in\mathbb R$ and $f$ is a smooth, compactly supported
function on $\mathbb R^2$ with $\int_{\mathbb R^2}f(u)du=0$, one has
\be
\label{gff}\media{e^{i\alpha\e^2\sum_{\bf x}h_{\bf x}f(\e{\bf
      x})}}\stackrel{\e\to0}\rightarrow e^{\tfrac{K\a^2}{4\p^2} \int f(u) f(v) \log|u-v|dudv},
\ee
where $\e^{-1}$ represents the coarse-grain scale, to be sent to infinity after the thermodynamic limit.}
\vskip.2truecm
The choice of the specific interaction in \eqref{1} is just for illustrative purposes: the same result remains valid for generic finite range interactions, translationally and rotationally invariant.

Eq.\eqref{gff} can be re-read in a more evocative form: if $\c_0$ is a smooth, compactly supported, probability distribution 
centered at the origin, $\c_\x(u)=\c_0(u-\x)$ is its translate, and $\tilde h_\e(\x)=\e^2\sum_\xx h_\xx \c_\x(\e \xx)$,
then choosing $f=\c_\x-\c_\h$ in \eqref{gff} we find
\be\lim_{\e\to 0}\media{e^{i\alpha (\tilde h_\e(\x)-\tilde h_\e(\h))
}}\ {\simeq}\ ({\rm const.})|\x-\h|^{-K\a^2/(2\p^2)},\nonumber
\ee
asymptotically as $|\x-\h|\to\infty$. The left side is the coarse-grained ``electric correlator": 
our theorem proves its anomalous power law decay at large distances.

An important step in proving the above result is the computation of the asymptotic behavior of the dimer correlation
by CRG methods \cite{M,GMT}. In the limit $L\to\infty$, 
if $m\neq 0$, it decays exponentially at large distances with rate $O(m^{1+\h_m(\l)})$ (gaseous phase), with $\h_m(\l)$ an analytic function such that $\h_m(0)=0$. If $m\to 0$, it decays as a power law (liquid phase): e.g., if  $b,b'$ are both horizontal
with $b'-b=(x_1,x_2)$ and $z=x_1+ix_2$, then
it decays polynomially, with critical exponent $\min\{2,2+\h(\l)\}$, and $\h(\l)=-(32/\p)\l+O(\l^2)$ an analytic function of $\l$,
\be\!\! \media{{\openone}_b;{\openone}_{b'}}\!=\!
\frac{(-1)^{x_1+x_2}}{2\p^2}{K(\l)}{\rm Re}\frac1{z^2}
+\small{(-1)^{x_1}}\frac{\bar K(\l)}{|z|^{2+\eta(\l)}}+h.o.\label{e3bis}\ee
Here $K(\l)$ is the same as in \eqref{e1.1}, $\bar K(\l)$ is another analytic function such that $\bar K(0)=1$, and $h.o.$ indicates faster decaying terms at infinity.
The above formula reduces as $\l\to0$ to the one known by Kasteleyn's exact solution. The main effect of the interaction is to produce an anomalous exponent in the
second term, in agreement with the numerical simulations of \cite{AJMPMT}. Remarkably,
there are no radiative corrections to the exponent of the first term.
The model belongs to the same universality class as the XXZ chain, vertex models and Luttinger liquids.

While the dimer characteristic function is a local observable, the height differences are non-local ``string" observables,  as apparent from  
(\ref{e31}). Even at $\l=0$, the computation of the height correlation is very subtle. Indeed, 
by inserting the $\l=0$ version of \eqref{e3bis} into \eqref{e31}, one gets an apparently very singular expression: take 
e.g. $\xx$ and $\yy$ on the same horizontal line. In the large separation limit,
the object of interest is formally proportional to 
$\int_{\x}^{\h}\!\! \frac{du dv}{(u-v)^2}$, where $\x,\h$ are the (suitably rescaled) horizontal  coordinates of $\xx$ and $\yy$. 
Such an integral requires a proper interpretation, because of its singularity at $u=v$, and the result depends on the specific ultraviolet regularization. Of course, an ``ad hoc" regularization can be chosen 
\cite{N} in order to reproduce the expected result, but the problem remains of a general 
derivation, which can unambiguously return the correct exponents without any external bias. The problem was finally 
solved in \cite{Ke}, and
the $(1/\pi^2)$ factor in front of the logarithm in \eqref{e1.1} at $\l=0$ was rigorously computed, 
by taking advantage of DH (lattice) methods.  
In the interacting case, the problem is much more puzzling. In fact, in addition to the problem of the ultraviolet divergences affecting the computation of the $(1/\pi^2)$ prefactor,
the anomalous decay in \eqref{e3bis}, once inserted into \eqref{e31}, may change the logarithmic growth into an anomalous growth. 
Our theorem proves that this is not the case: logarithmic fluctuations are robust, stability being guaranteed by sophisticated cancellations arising from emerging chiral symmetry. 
Spurious ultraviolet divergences are avoided by using the irrelevant terms coming from the lattice: in this respect, 
the use of exact CRG methods (which, in contrast to field theoretic RG, takes the irrelevant terms into full account) is essential.
A detailed proof of  our main theorem  is rather technical and
is given elsewhere \cite{GMT}; below we explain its main ideas.

{\it Sketch of the proof.}
The first step consists in an exact rewriting of the finite volume/finite lattice {\it generating function} of dimer correlations, $\mathcal Z(A)$, (defined so that $\media{\openone_{b_1};\cdots ;\openone_{b_k}}=\frac{\partial^k}{\partial A_{b_1}\cdots \partial A_{b_k}} \log \mathcal Z(A))|_{A=0}$, $b_i$ labeling the nearest neighbor bonds) 
as a finite Grassmann integral
\cite[Section 2]{GMT}:
\be
 \mathcal Z(A)=\frac12\sum_{\theta,\tau}C_{\theta,\tau}\int _{\th,\t}P_{\th,\t}(d\psi)e^{V(\psi)+B(\psi, A)}.
\label{sa}\ee
Here $\psi_\xx$ are {\it Grassmann variables}, 
$V$ is sum of monomials in $\psi_\xx$ of order 4 or higher, 
$B(\psi, A)$ is a source term, sum of monomials in $\psi$ and in $A$,
$\th/\t\in\{0,1\}$ label the boundary conditions for the Grassmann variables in the horizontal/vertical directions
($0/1$ corresponding to periodic/antiperiodic conditions), 
and $C_{0,0}=-1$, while $C_{\th,\t}=+1$ otherwise. By cluster expansion methods, we prove that $V$ and $B$ are analytic in $\l$. $P_{\th,\t}(d\psi)$ is a gaussian Grassmann integration with propagator $g(\xx,\yy)$ 
\be
\frac{1}{L^2}\sum_{\kk}
{e^{-i \kk(\xx-\yy)}}\frac{i\sin k_1+\sin k_2+m(-1)^{y_1}\cos k_1}{
2D(\kk,m)
}\label{prop}
\;,\ee
where $D(\kk,m)=m^2+(1-m^2)(\sin k_1)^2+(\sin k_2)^2$, and $k_1,k_2$ are in $(2\p/L)\mathbb Z$ or 
$(2\p/L)(\mathbb Z+1/2)$, depending on boundary conditions.

If $\l=0$, then $V=0$, in which case the integral is gaussian and can be computed exactly. When $\l\not=0$ the integral is not gaussian, and it can be evaluated by a multiscale analysis using CRG methods \cite{M}. 
We are interested in the case of $m$ small or vanishing. As $L\to\infty$ and $m\to 0$, the propagator in \eqref{prop} 
becomes singular in correspondence
of four momenta, namely  ${\bf p}_{1}=(0,0)$, ${\bf p}_{2}=(\pi,0)$, ${\bf p}_{3}=(\p,\p)$, ${\bf p}_{4}=(0,\p)$. Therefore, $g(\xx,\yy)$ can be naturally written as the
superposition of four terms, each of which is concentrated in momentum space around one of the singularities. 
Correspondingly, we decompose the Grassmann field as:
\be \psi_\xx=e^{i{\bf p}_1 \xx}\psi_{\xx,1}-i e^{i{\bf p}_2 \xx}\psi_{\xx,2}
+ie^{i{\bf p}_3 \xx}\psi_{\xx,3}+e^{i{\bf p}_4 \xx}\psi_{\xx,4},\label{h}\ee
where $\psi_{\xx,\g}$ are Grassmann variables,
often referred to as Majorana variables, since their effective action is 
a lattice regularization of the standard 2D Majorana action. Their propagator is block-diagonal, the fields with $\g=1,2$ being independent of $\g=3,4$; the 
propagator $G(\xx-\yy)$ of the $\g=1,2$ fields is the same as that of the $\g=3,4$ fields, and reads (using the symbol $\int d\kk/(2\p)^2$ as a shorthand for the discrete sum in \eqref{prop}): $G(\xx)=$
\be\frac1{Z}
\int\frac{d\kk}{(2\pi)^2}\frac{\chi(\kk) e^{-i\kk\xx}}{2D(\kk,m)}
\begin{pmatrix}
    i \sin k_1+\sin k_2 & i m\cos k_1\\
-im\cos k_1 & i \sin k_1-\sin k_2
  \end{pmatrix}\nonumber
\ee
where $\chi(\kk)$ is a smoothed characteristic function of the set $\max\{|k_1|, |k_2|\}\le \p/2$, 
and $Z=1$.
 To evaluate 
the Grassmann integral (\ref{sa}) we use \eqref{h} and 
write the propagator $G(\xx)$ as sum of propagators living on
momentum scales $2^h,h\le 0$.  
After integrating the scales $0,\dots,h+1$, the $(\th,\t)$ contribution to $\mathcal Z(A)$ 
is rewritten as
$$ e^{S_h(A)}
\int _{\th,\t} P_{Z_h,m_h}(d\psi^{(\le h)})e^{V^{(h)}(\sqrt{Z_h}\psi^{(\le h)})+B^{(h)}(\sqrt{Z_h}\psi^{(\le h)}, A)}
$$
where $P_{Z_h,m_h}$ has propagator $G^{(h)}(\xx)$, defined in the same way as $G(\xx)$, with $Z$ replaced by $Z_h$, 
$m$ by $m_h$ and $\c(\kk)$ by $\c_h(\kk)$, a (smoothed) characteristic function of the set $|\kk|\le (\p/2)2^h$.
The effective potential $V^{(h)}$ is:
 $$%
V^{(h)}(\psi)=\l_h\sum_\xx \psi_{\xx,1}\psi_{\xx,2}\psi_{\xx,3}\psi_{\xx,4}+ir., $$
where $ir.$ indicates the irrelevant terms (non-local quartic terms, and terms of order $6$ or higher in $\psi$). Remarkably, the kernels of the irrelevant terms in $V^{(h)}$ are {\it analytic} in $\l$ provided that 
$|Z_{h+1}/Z_h-1|,|\l_h|$ are sufficiently small, as long as 
$|m_h|<2^h$: the proof of this fact uses fermionic cluster expansion methods, including the use of Gram-Hadamard 
determinant bounds. Similarly, under the same assumptions, the effective  source $B^{(h)}$ is analytic in $\l$. Its  
structure is expressed most easily by using Dirac rather than Majorana fields: the former are 
defined as 
$\psi^\pm_{\xx,1}:=\frac1{\sqrt2}(\psi_{\xx,1}\mp i\psi_{\xx,3})$, $\psi^\pm_{\xx,-1}:=\pm\frac i{\sqrt2}(\psi_{\xx,2}\mp i\psi_{\xx,4})$, 
and they are referred to as Dirac variables, because their action is the lattice analogue of that of 2D Dirac fields. 
In terms $\psi^\pm_\o$, the effective source reads:
$$ B^{(h)}(\psi)=\frac{Z^{(1)}_{h}}{Z_h}F_1(\psi,{\bf
  J})+\frac{Z^{(2)}_{h}}{Z_h}F_2(\psi,{\bf J})+ir.\;,$$
where $ir.$ are the irrelevant terms (non local, or of higher order in $A$ or $\psi$ as compared to $A\psi\psi$). Moreover, 
denoting $J_{\xx,i}=J_{(\xx,\xx+\hat e_i)}$ with $J_b=e^{A_b}-1$: 
$F_1=2\sum_{\xx,\ \o=\pm}(-1)^{\xx}(J_{\xx,1}+i\o J_{\xx,2})\psi^+_{\xx,\o}\psi^-_{\xx,\o}$, and 
$F_2=2\sum_{\xx,\ \o=\pm}\big[(-1)^{x_1}J_{\xx,1}+i\o(-1)^{x_2}J_{\xx,2}\big]\psi^+_{\xx,\o}\psi^-_{\xx,-\o}$.
Summarizing, the effective theory on scale $h$ has the same structure as a theory of interacting 2D lattice Dirac fermions
with 
a wave function renormalization $Z_h$, an effective mass $m_h$, an effective coupling $\l_h$, and effective source 
couplings $Z^{(1)}_h, Z^{(2)}_h$. It is completely analogous to that obtained in the multiscale analysis of the 8 Vertex, Ashkin-Teller, XXZ, or Luttinger liquid models \cite{M}: the only differences have to be found 
in the oscillating factors appearing in the definition of $F_1,F_2$ and in the specific structure of the irrelevant terms.
The flow equation for the effective couplings of all these models is the same, up to irrelevant contributions, which are 
exponentially negligible in the infrared limit. Therefore, $\l_h$ approaches exponentially, as $h\to-\infty$, a line of fixed points: $\l_{-\infty}(\l)=-32\l(1+O(\l))$. Moreover, $Z_h\sim 2^{\h(\l)h}$, $Z_h^{(i)}\sim 2^{\h_i(\l)h}$, $m_h
\sim m\, 2^{\h_m(\l) h}$,
where $\sim$ means that the ratio of the two sides is bounded from above and below by two universal positive 
constants, uniformly in $h$. Remarkably, using the emergent chiral gauge symmetry of the theory, 
we find that $\h=\h_1$, which implies the robustness (exact non-renormalization) of the exponent 2 in the first term of 
\eqref{e3bis}. The integration goes on until $m_h\simeq 2^h$, at which point the Dirac field is massive and can be integrated in one step. If $m\to0$ and $L\to\infty$, the integration has no infrared cutoff.

\begin{figure}[ht]
\includegraphics[width=.25\textwidth]{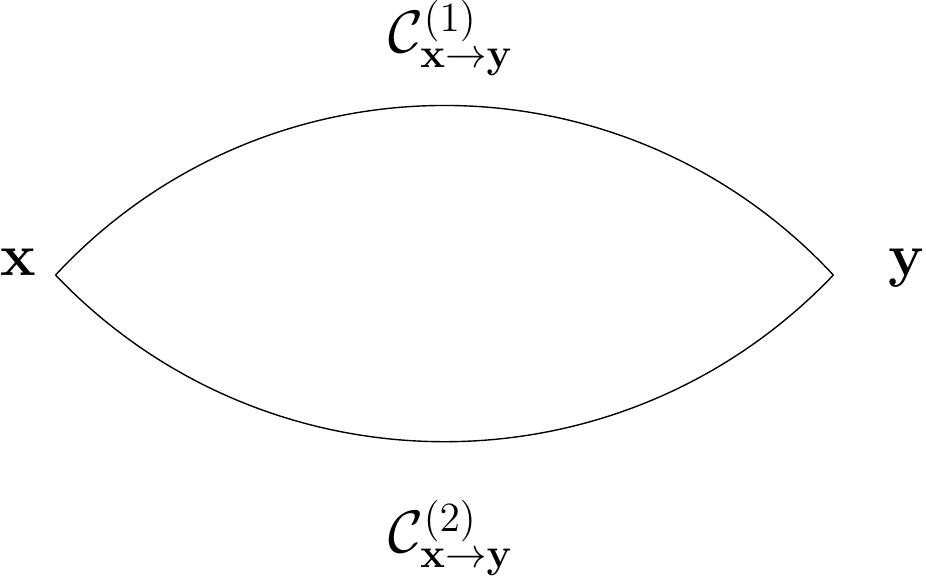}
\caption{A schematic view of the paths along which $b_1,b_2$ are summed over, to be called ${\mathcal C}^{(1)}_{\xx\to\yy}$ and ${\mathcal C}^{(2)}_{\xx\to\yy}$.}
\label{figcammini}
\end{figure}

In order to evaluate the height fluctuations, we use the path-independence of the height difference, which is a (weak) instance of DH. We proceed as in \cite{Ke}(c). Consider e.g. the height variance: in the right side of \eqref{e31}
we deform the two paths along which $b_1$ and $b_2$ are summed over, in such a way that they are ``as much separated as possible", as in Fig.\ref{figcammini}. 
In the vicinity of $\xx$ and $\yy$, the two paths are lattice approximations of straight lines, departing from and arriving at the points $\xx,\yy$ in {\it different} directions. 
After the path deformation, we replace the dimer correlation in the right side of \eqref{e31} by its asymptotic expression \eqref{e3bis} (and its analogues in the cases that $b,b'$ have different orientations). The $h.o.$ terms contribute a finite constant, uniformly in $|\xx-\yy|$. 
The contribution to  \eqref{e31} from the term with decay exponent 2 (let us call it $Cont_2$) reads: $Cont_2=-\frac{K(\l)}{2\pi^2}\sum_{b_1\in \mathcal C^{(1)}_{\xx\to \yy}}\sum_{b_2\in \mathcal C^{(2)}_{\xx\to \yy}}{\rm Re}\frac{\D z_{b_1}\D z_{b_2}}{(z_{b_1}-z_{b_2})^2}$,
where $z_{b_i}$ are the representatives in complex coordinates of the centers of the bonds $b_i$, and $\D z_{b_i}$ are the oriented elementary path elements of $\mathcal C^{(i)}_{\xx\to\yy}$ crossing $b_i$, expressed in complex coordinates. Note that no oscillatory factor appears in $Cont_2$: the factors $\s_{b_1}\s_{b_2}$ in \eqref{e31}
compensate exactly the oscillatory factor of  the term under consideration in the dimer correlation.
$Cont_2$ is the Riemann approximation to $-\frac{K(\l)}{2\pi^2}{\rm
  Re}\int_{\g_1}dz\int_{\g_2}dw\frac1{(z-w)^2}$,
where $\g_1$ and $\g_2$ are two {\it completely disjoint} complex paths (this is what makes the integral non-singular!) 
going from $z_\xx=z_{b^{(1)}_\xx}$ to $z_\yy=z_{b^{(1)}_\yy}$,
and from $z_\xx'=z_{b^{(2)}_\xx}$ to $z_\yy'=z_{b^{(2)}_\yy}$, where $b^{(i)}_\xx$ and $b^{(i)}_\yy$ are the first and last bonds of $\mathcal C^{(i)}_{\xx\to\yy}$.
Its value is $\frac{K}{2\p^2}{\rm Re}\log\frac{(z_{\yy}'-z_{\xx})(z_{\xx}'-z_{\yy})}{(z_{\yy}'-z_{\yy})(z_{\xx}'-z_{\xx})}$,
which is the same as \eqref{e1.1} up to a bounded error. 
Finally, consider the contribution to \eqref{e31} from the term with exponent $2+\h$:  
in this case the factors $\s_{b_1}\s_{b_2}$ do not compensate exactly with the oscillatory signs in the dimer correlation;
the left-over oscillations act, after summation along the paths, as discrete derivative, which effectively makes this term decay faster, thus making  its contribution to \eqref{e31} finite, uniformly in $|\xx-\yy|$. Similar considerations apply to higher order cumulants, and (\ref{gff}) follows as a corollary.

In conclusion, we presented a rigorous microscopic derivation of massless gaussian free field behavior
of the height field of a non integrable interacting dimer model. Our method combines constructive field theory techniques 
with discrete holomorphicity ideas, which are used  for the first time in a unified way to analyze a non-local fermionic 
observable. The method can be applied to several other non-integrable 2D critical theories
and we expect it to be capable, in perspective, of rigorously proving conformal invariance of the scaling limit. 

{\bf Acknowledgments.} This research was supported by the  ERC Starting Grant CoMBoS (g.a. n$^o$ 239694; A.G. and V.M.) and the Marie Curie Fellowship
DMCP (F.T.).

\bibliographystyle{amsalpha}

\end{document}